\documentclass[journal]{IEEEtran}

\ifCLASSINFOpdf
\else
   \usepackage[dvips]{graphicx}
\fi
\usepackage{url}

\usepackage{graphicx}
\usepackage{cite}
\usepackage[dvipsnames]{xcolor}
\usepackage{multirow}
\usepackage{hyperref}
\usepackage{soul} 
\usepackage{color,xcolor} 
\soulregister{\cite}7 
\soulregister\ref7
\begin{document}

\title{Text-guided HuBERT: Self-Supervised Speech Pre-training via Generative Adversarial Networks}

%

\author{Duo Ma, Xianghu Yue, Junyi Ao, Xiaoxue Gao, Haizhou Li, \IEEEmembership{Fellow, IEEE}



\thanks{This work was supported in part by Huawei Noah’s Ark Lab, and in part by the National Natural Science Foundation of China (Grant No. 62271432); Shenzhen Science and Technology Program ZDSYS20230626091302006; and Shenzhen Science and Technology Research Fund (Fundamental Research Key Project Grant No. JCYJ20220818103001002). (\textit{Corresponding author: Xianghu Yue.})}

\thanks{Duo Ma, Junyi Ao, and Haizhou Li, Shenzhen Research Institute of Big Data, School of Data Science, The Chinese University of Hong Kong, Shenzhen 518172, China (email: maduo@cuhk.edu.cn; junyiao1@link.cuhk.edu.cn; haizhouli@cuhk.edu.cn).}
\thanks{Xianghu Yue, Xiaoxue Gao, Department of Electrical and Computer and Engineering, National University of Singapore, Singapore 119077 (e-mail:
xianghu.yue@u.nus.edu, xiaoxue.gao@u.nus.edu).}
\vspace{-0.1cm}
}

\markboth{Journal of \LaTeX\ Class Files, Vol. 14, No. 8, August 2015}
{Shell \MakeLowercase{\textit{et al.}}: Bare Demo of IEEEtran.cls for IEEE Journals}
\maketitle

\begin{abstract}
Human language can be expressed in either written or spoken form, i.e. text or speech. 
Humans can acquire knowledge from text to improve speaking and listening. However, the quest for speech pre-trained models to leverage unpaired text has just started.
In this paper, we investigate a new way to pre-train such a joint speech-text model to learn enhanced speech representations and benefit various speech-related downstream tasks.
Specifically, we propose a novel pre-training method, text-guided HuBERT, or T-HuBERT, which performs self-supervised learning over speech to derive phoneme-like discrete representations. 
And these phoneme-like pseudo-label sequences are firstly derived from speech via the generative adversarial networks (GAN) to be statistically similar to those from additional unpaired textual data.
In this way, we build a bridge between unpaired speech and text in a unsupervised manner.
Extensive experiments demonstrate the significant superiority of our proposed method over various strong baselines, which achieves up to 15.3\% relative Word Error Rate (WER) reduction on the LibriSpeech dataset.
\end{abstract}


\IEEEpeerreviewmaketitle

\section{Introduction}
\IEEEPARstart Speech pre-trained models~\cite{baevski2020wav2vec, hsu2021hubert, chen2022wavlm, pmpc2022} have recently advanced the state-of-the-art in speech-related tasks, such as speech recognition~\cite{chen2022wavlm, baevski2020wav2vec, hsu2021hubert}, speaker verification~\cite{speaker1, speaker2} and speech enhancement~\cite{se1, se2}.
Speech and text are two important mediums of human communication, and they can be directly converted into each other through speech synthesis and recognition systems.
We human beings can acquire knowledge from additional text to improve speaking and listening.
However, the quest for speech pre-trained models to leverage unpaired text has just started.
SpeechLM~\cite{zhang2022speechlm}, MAESTRO~\cite{chen2023maestro}, and SLAM~\cite{bapna2021slam}, utilize paired data to bridge the gap between speech and text and learn enhanced speech representations.
token2vec~\cite{yue2023token2vec} offers another approach by converting speech waveforms into discrete code sequences and text into phoneme sequences, thus leveraging unpaired textual data to enhance speech pre-training. 
Despite the conceptual similarity between speech codes and text phonemes, token2vec does not explicitly establish a direct correspondence between the two modalities.

In this work, we propose a novel technique to leverage additional unpaired text for speech pre-training.
By enhancing the phonetic information of speech representations, we finally benefit various downstream speech-related tasks.
To achieve this, we build a bridge between unpaired speech and text in an unsupervised manner.
Specifically, we derive phoneme sequences from speech data via the generative adversarial networks (GAN) to be statistically similar to those from the text data.
These derived phoneme-like pseudo labels then serve together with the original targets as hierarchical targets for speech pretraining to learn high-level semantic speech representations, following the principles of HuBERT~\cite{hsu2021hubert}
The contributions of T-HuBERT can be summarized as follows:
1) We first leverage unpaired text information to obtain phoneme-like modelling units from speech via unsupervised training, thus eliminating the need of paired data as a bridge for joint pre-training of speech and text~\cite{zhang2022speechlm, zhang2022speechut, bapna2021slam, chen2022maestro}.
2) Compared with the hidden unit of HuBERT, the derived phoneme-like modelling units could effectively guide the pre-trained model to improve its speech representation. 
3) Based on previous empirical evidence in~\cite{pasad2023comparative}, we show that the phoneme-like modelling units and the hidden units can jointly guide the pre-trained model to achieve better speech representation.

\begin{figure*}[htbp]
\centering
\includegraphics[width=0.85\textwidth]{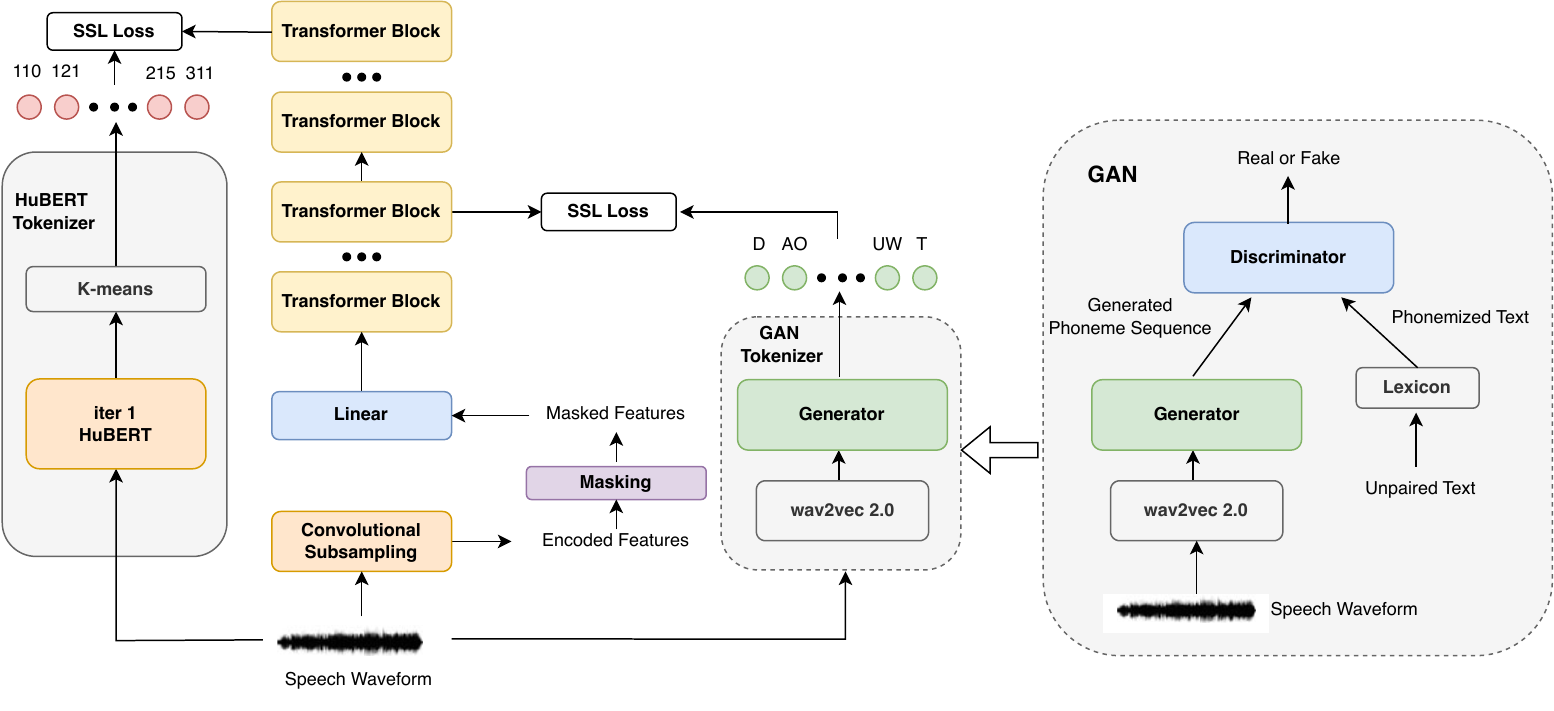}
\vspace{-0.2cm}
\caption{
Illustration of our T-HuBERT model.
In the left part, we use HuBERT to discretize the speech to get the pseudo-code labels, and the middle part is the main structure of our model.
In the right part, we use the GAN network to discretize the speech to get the phoneme-like pseudo-label sequences.}
\label{fig:voicelm}  
\vspace{-0.25cm}
\end{figure*}

\section{Methodology}
\label{sec:approach}
Given unpaired speech and text data, how does the speech pre-trained model leverage the additional unpaired text to benefit the downstream speech-related tasks? 
As speech and text are two different forms of a language, they are inherently similar in terms of phonetic, phonotactic, and lexical distribution. 
The question is how we draw an equivalence between the self-organized phonetic units, i.e. HuBERT codes, and their phonetic equivalents derived from text. 

The success of HuBERT shows that a poor teacher can teach a good student, so it is logical that if a better teacher is found, a better student speech pre-training model can be delivered.
To this end, inspired by wav2vec-U 2.0~\cite{liu2023towards}, we introduce an unsupervised approach to obtain the discrete representations of frame-level speech waveform via the adversarial training between speech and text.
In this way, we build a bridge between unpaired speech and text without an explicit supervision signal.
Subsequently, these obtained phoneme-like sequences serve as additional self-supervised targets on the intermediate layer during pre-training, thus enhancing the phonetic information of the learned speech representations.
The overall framework of our proposed method, T-HuBERT, is depicted in Figure \ref{fig:voicelm}.

\subsection{HuBERT-based Tokenizer}
\label{ssec:hidden_unit}
We follow the main settings of HuBERT, in which there are two iterative re-clustering and re-training stages.
In the first stage, the discrete targets are obtained by clustering on the MFCC features.
In the second stage, a new generation of training targets is derived by clustering on the learned representations from the first stage model.
For the pre-training of T-HuBERT, we adopt a similar two-iteration strategy.
Specifically, we apply these k-means codes, obtained using the HuBERT model from the first iteration, as the targets of the top layer of the T-HuBERT model.

\vspace{-0.15cm}
\subsection{GAN-based Tokenizer}
\label{ssec:phoneme}
To leverage additional text data and improve the phonetic information of the pre-training, we utilize the GANs~\cite{gans} as a bridge between unpaired speech and text, to discretize the speech waveform into phoneme-like pseudo-label sequences in an unsupervised manner.
Specifically, as shown in the right part of Figure~\ref{fig:voicelm}, the generator consumes speech, while the input of the discriminator is the real phoneme sequences of unpaired text.
During adversarial training, the discriminator is trained to distinguish the output of the generator and real phoneme sequences.
The goal of the generator is to output phoneme sequences that cannot be distinguished by the discriminator.

We adopt the same unsupervised training strategy as wav2vec-U 2.0~\cite{liu2023towards}.
Specifically, we train the network, on a traditional GAN loss, $\mathcal{L}_{gan}$, a gradient penalty term, $\mathcal{L}_{gp}$, for better convergence, a smoothness penalty term, $\mathcal{L}_{sp}$, to encourage consecutive speech frames to output the same phonemes, a phoneme diversity term, $\mathcal{L}_{pd}$, to diverse phoneme usage, and a self-supervised loss $\mathcal{L}_{ss}$, to encourage the generated phonemes to match the input speech.
The final GAN training objective is formulated following:
\begin{equation}
    \mathcal{L} = \min_{\mathcal{G}} \max_{\mathcal{C}} [\mathcal{L}_{gan} + \lambda \mathcal{L}_{gp} + \gamma \mathcal{L}_{sp} + \eta \mathcal{L}_{pd} + \delta \mathcal{L}_{ss} ]
\end{equation}

After adversarial training, we use the generator to discretize the speech waveform into frame-level phoneme-like pseudo sequences, which can be considered to be obtained under the supervision of additional unpaired text.

\vspace{-0.25cm}
\subsection{Multi-Unit and Multi-Layer Supervision Objective}
\label{ssec:mm}
Our model architecture design mainly follows the HuBERT-base model, consisting of a convolutional feature encoder and a transformer encoder.
The convolutional encoder is composed of seven temporal convolution blocks, which have 512 channels with strides $(5,2,2,2,2,2,2)$ and kernel widths $(10,3,3,3,3,2,2)$, resulting in a 20ms frame rate for audio sampled at 16kHz.
The transformer encoder contains 12 blocks, with model dimension 768, feed-forward dimension 3072, and 12 attention heads.
Before these features are fed to the transformer encoder, we randomly mask them in time dimension with a probability of 0.8.
For the transformer encoder, instead of the simple relative position embedding used in HuBERT, we use a bucket relative position embedding~\cite{Raffel_Shazeer_Roberts_Lee_Narang_Matena_Zhou_Li_Liu_2019}, which generates different learned embeddings based on the offset between the ``key" and the ``query" in the vanilla self-attention. 

To leverage the HuBERT-based and GAN-based targets, we further propose a multi-unit and multi-layer supervision objective, which could effectively guide the pre-trained model to learn rich phonetic information, thus benefiting downstream tasks.
Specifically, during pre-training, we further select one intermediate layer $k$ as the supervised layer of GAN-based targets, besides the traditional HuBERT-based targets for the top layers, 
and calculate the masked prediction loss, which can be
summarized as follows:
\begin{equation}\label{eq:2}
    \mathcal{L} = -\sum_{l \in \{k, L\}}\sum_{t\in \mathcal{M}}{log \ p^l(c^l_t|{\mathbf{o}^l_t)}}
\end{equation}
where $L$ is the number of encoder layer, $\mathcal{M}$ denotes the set of masked timesteps, $c^l_t$ denotes the corresponding codes at timestep $t$, and $\mathbf{o}^l_t$ denotes the output hidden states of layer $l$ at timestep $t$.
The distribution over the codes is parameterized with:
\begin{equation}
    {p^l(c^l_t|{o^l_t)} = \frac{exp(sim(\mathbf{W}^l{o^l_t}, \mathbf{e}^l(c^l_t))/\tau)}{\sum_{c^\prime\in C}{exp(sim(\mathbf{W}^l{o^l_t}, \mathbf{e}^l(c^\prime))/\tau})}}  
\end{equation}
where $\mathbf{W}$ is the project matrix, and $\mathbf{e}$ is the embedding for code $c$.




\begin{table}
\centering
\caption{WER results on LibriSpeech test sets when fine-tuned on 1 hour, 10 hours, the train-clean-100, train-clean-360, and train-other-500 splits of LibriSpeech.}
\scalebox{0.9}{
\begin{tabular}{c c c c c c c } 
\hline\hline
\multirow{2}{*}{Model} &\multicolumn{2}{c}{No LM} & \multicolumn{2}{c}{LM}  \\
& clean & other &  clean & other   \\
\hline
\textbf{1-hour subset}\\
\hline
DeCoAR 2.0~\cite{wang2021self} & - & - & 13.8 & 29.1  \\
DiscreteBERT~\cite{baevski2019effectiveness}  & - & - & 9.0 &17.6\\
wav2vec 2.0~\cite{baevski2020wav2vec}  & 24.5 & 29.7  & 5.5 &11.3 \\
HuBERT~\cite{hsu2021hubert}  & 20.9 & 27.5 & 6.1 &11.3 \\
WavLM~\cite{chen2022wavlm}  & 24.5 &29.2 & 5.7 &10.8 \\
token2vec~\cite{yue2023token2vec}&20.0&25.8&-&-\\
\hline
T-HuBERT           & 16.4& 22.2 &6.4&11.9 \\
\hline\hline
\textbf{10-hour subset}\\
\hline
DeCoAR 2.0~\cite{wang2021self}  & - & - & 5.4 & 13.3  \\
DiscreteBERT~\cite{baevski2019effectiveness}  & - & - & 5.9 &14.1\\
wav2vec 2.0~\cite{baevski2020wav2vec}  & 11.1 & 17.6  & 4.3 &9.5 \\
HuBERT~\cite{hsu2021hubert}  & 10.1 & 16.8 & 4.3 &9.4 \\
WavLM~\cite{chen2022wavlm}  & 9.8 &16.0 & 4.3&9.2 \\
token2vec~\cite{yue2023token2vec} &9.0&14.9&-&-\\     
\hline
T-HuBERT            & 7.8&13.7 &4.1&9.1\\
\hline\hline
\textbf{100-hour subset} \\
\hline
DeCoAR 2.0~\cite{wang2021self}  & - & - & 5.0 & 12.1  \\
DiscreteBERT~\cite{baevski2019effectiveness}  & - & - & 4.5 &12.1\\
wav2vec 2.0~\cite{baevski2020wav2vec} & 6.1 & 13.3  & 3.4 &8.0 \\
HuBERT~\cite{hsu2021hubert} & 6.3 & 13.2 & 3.4 &8.1 \\
WavLM~\cite{chen2022wavlm}  & 5.7 &12.0 & 3.4&7.7 \\
token2vec~\cite{yue2023token2vec} &5.1&11.8&-&-\\     
PBERT~\cite{wang2022supervision} &4.2 &9.5&3.1&7.2\\
SpeechLM-P~\cite{zhang2022speechlm}&3.4 &8.1 &2.7 &6.2\\
\hline
T-HuBERT           & 4.8& 10.0 &3.1&7.0 \\
\hline\hline
\textbf{360-hour subset}\\
\hline
HuBERT & 4.8 & 11.0 &3.0 & 7.2\\
T-HuBERT & 3.8 & 9.1 &2.7 &6.4\\
\hline\hline
\textbf{500-hour subset}\\
\hline
HuBERT & 4.8 & 10.1 & 3.0 & 6.7\\
T-HuBERT &3.7 &8.1  & 2.6 &5.9\\
\hline\hline
\end{tabular}}
\label{tab_main}
\vspace{-0.4cm}
\end{table}



\vspace{-0.3cm}
\section{Experiments}
\label{sec:exp}
\vspace{-0.15cm}
\subsection{Datasets}
\label{ssec:data}
For unsupervised speech pre-training, we use the publicly available 960-hour LibriSpeech~\cite{Panayotov_Chen_Povey_Khudanpur_2015} dataset.
For unpaired text data, we use the LibriSpeech LM corpus~\footnote{https://www.openslr.org/resources/11/librispeech-lm-norm.txt.gz}, containing about 40M English sentences.
For the downstream ASR task, we use the \textit{train-clean-100} split of LibriSpeech, the 1-hour and 10-hour split data of Libri-Light~\cite{kahn2020libri} as the low-resource scenarios.
We further evaluate on the SUPERB benchmark~\cite{yang2021superb}, which is designed to provide a standard and comprehensive testbed for pre-trained speech models on various speech tasks.


\vspace{-0.15cm}
\subsection{Experimental Setup}
\label{ssec:exp_setup}
All pre-training experiments are conducted using fairseq toolkit ~\cite{ott2019fairseq}.
For GAN-based tokenizer, we mainly follow the training process and hyper-parameters in \cite{liu2023towards}.
Specifically, we first use the unsupervised voice activity detection~\cite{tan2020rvad} to remove the silences from speech and then feed the speech into the pre-trained speech model, i.e.,  wav2vec 2.0-large\footnote{https://dl.fbaipublicfiles.com/fairseq/wav2vec/wav2vec\_vox\_new.pt}, to extract high-level speech representations.
For unpaired text, we convert the word sequences into phoneme sequences using an off-the-shelf phonemizer~\cite{g2pE2019}, which contains a total of 41 monophones, including a silence symbol.
The generator consists of a single-layer CNN with batch normalization and the discriminator employs a three-layer CNN that takes either the output of the generator or a one-hot representation of the real phoneme sequence as input.

All pre-training models are trained on 8 V100 32GB GPUs with a gradient accumulation number of 4 for a total of 400k steps.
We randomly sample starting positions with a probability of 0.08 and apply a mask to the subsequent 10 time steps.
We employ the Adam optimizer~\cite{adam} with a weight decay of 0.01 and $\beta = (0.9, 0.98)$. 
The learning rate follows a linear increase for the first 32k steps and subsequently decays linearly to 0. The peaking learning rate is 5e-4.
Following~\cite{hsu2021hubert, baevski2020wav2vec}, we use the Connectionist Temporal Classification (CTC) \cite{graves2006connectionist} loss for downstream ASR fine-tuning.
The target vocabulary includes 26 English characters, a space token, and an apostrophe.
We train the model with Adam optimizer and a tri-stage rate schedule where the learning rate is warmed up for the first 10\% of the updates, held constant for the next 40\%, and then linearly decayed for the remainder.
During the inference stage, we use the flashlight \cite{kahn2022flashlight} \footnote{https://github.com/flashlight/text} beam search decoder with beam size 1500 for a 4-gram language model (LM)~\footnote{https://www.openslr.org/resources/11/4\-gram.arpa.gz}.

\begin{table*}\footnotesize
\centering
\caption{Universal speech representation evaluation on the SUPERB benchmark.}
\scalebox{0.85}{
\begin{tabular}{c|c|c|ccc|cccc|cc|c}
\hline\hline
\multicolumn{1}{c|}{\multirow{3}{*}{Method}} & \multirow{3}{*}{\#Params} & \multirow{3}{*}{Corpus} & \multicolumn{3}{c|}{Speaker} & \multicolumn{4}{c|}{Content} & \multicolumn{2}{c|}{Semantics} & \multicolumn{1}{c}{ParaL} \\ \cline{4-13} 
\multicolumn{1}{c|}{} &  &  & \multicolumn{1}{c|}{\multirow{2}{*}{\begin{tabular}[c]{@{}c@{}}SID\\ Acc$\uparrow$\end{tabular}}} & \multicolumn{1}{c|}{\multirow{2}{*}{\begin{tabular}[c]{@{}c@{}}ASV\\ EER$\downarrow$\end{tabular}}} & \multirow{2}{*}{\begin{tabular}[c]{@{}c@{}}SD\\ DER$\downarrow$\end{tabular}} & \multicolumn{1}{c|}{\multirow{2}{*}{\begin{tabular}[c]{@{}c@{}}PR\\ PER$\downarrow$\end{tabular}}} & \multicolumn{1}{c|}{\multirow{2}{*}{\begin{tabular}[c]{@{}c@{}}ASR\\ WER$\downarrow$\end{tabular}}} & \multicolumn{1}{c|}{\multirow{2}{*}{\begin{tabular}[c]{@{}c@{}}KS\\ Acc$\uparrow$\end{tabular}}} & \multirow{2}{*}{\begin{tabular}[c]{@{}c@{}}QbE\\ MTWV$\uparrow$\end{tabular}} & \multicolumn{1}{c|}{\multirow{2}{*}{\begin{tabular}[c]{@{}c@{}}IC\\ Acc$\uparrow$\end{tabular}}} & \multirow{2}{*}{\begin{tabular}[c]{@{}c@{}}SF\\ F1$\uparrow$\hspace{3.0mm} CER$\downarrow$\end{tabular}} & \multirow{2}{*}{\begin{tabular}[c]{@{}c@{}}ER\\ Acc$\uparrow$\end{tabular}} \\
\multicolumn{1}{c|}{} &  &  & \multicolumn{1}{c|}{} & \multicolumn{1}{c|}{} &  & \multicolumn{1}{c|}{} & \multicolumn{1}{c|}{} & \multicolumn{1}{c|}{} &  & \multicolumn{1}{c|}{} &  &  \\ \hline
TERA & 21.33M & LS960hr & \multicolumn{1}{c|}{57.57} & \multicolumn{1}{c|}{15.89} & 9.96 & \multicolumn{1}{c|}{49.17} & \multicolumn{1}{c|}{18.17} & \multicolumn{1}{c|}{89.48} & 0.0013 & \multicolumn{1}{c|}{58.42} & 67.50\hspace{3.0mm}54.17 & 56.27 \\ 
DeCoAR 2.0 & 89.84M & LS960hr & \multicolumn{1}{c|}{74.42} & \multicolumn{1}{c|}{7.16} & 6.59 & \multicolumn{1}{c|}{14.93} & \multicolumn{1}{c|}{13.02} & \multicolumn{1}{c|}{94.48} & 0.0406 & \multicolumn{1}{c|}{90.80} & 83.28\hspace{3.0mm}34.73 & 62.47 \\ 
wav2vec & 32.54M & LS960hr & \multicolumn{1}{c|}{56.56} & \multicolumn{1}{c|}{7.99} & 9.90 & \multicolumn{1}{c|}{31.58} & \multicolumn{1}{c|}{15.86} & \multicolumn{1}{c|}{95.59} & 0.0485 & \multicolumn{1}{c|}{84.92} & 76.37\hspace{3.0mm}43.71 & 59.79 \\ 
vq-wav2vec & 34.15M & LS960hr & \multicolumn{1}{c|}{38.80} & \multicolumn{1}{c|}{10.38} & 9.93 & \multicolumn{1}{c|}{33.48} & \multicolumn{1}{c|}{17.71} & \multicolumn{1}{c|}{93.38} & 0.0410 & \multicolumn{1}{c|}{85.68} & 77.68\hspace{3.0mm}41.54 & 58.24 \\ 
wav2vec 2.0 & 95.04M & LS960hr & \multicolumn{1}{c|}{75.18} & \multicolumn{1}{c|}{6.02} & 6.08 & \multicolumn{1}{c|}{5.74} & \multicolumn{1}{c|}{6.43} & \multicolumn{1}{c|}{96.23} & 0.0233 & \multicolumn{1}{c|}{92.35} & 88.30\hspace{3.0mm}24.77 & 63.43 \\ 
HuBERT & 94.68M & LS960hr & \multicolumn{1}{c|}{81.42} & \multicolumn{1}{c|}{5.11} & 5.88 & \multicolumn{1}{c|}{5.41} & \multicolumn{1}{c|}{6.42} & \multicolumn{1}{c|}{96.30} & 0.0736 & \multicolumn{1}{c|}{98.34} & 88.53\hspace{3.0mm}25.20 & 64.92 \\ 
SpeechLM-P & 94.70M & LS960hr & \multicolumn{1}{c|}{75.24} & \multicolumn{1}{c|}{5.97} & 7.34 & \multicolumn{1}{c|}{3.10} & \multicolumn{1}{c|}{4.98} & \multicolumn{1}{c|}{94.09} & 0.0410 & \multicolumn{1}{c|}{97.68} & 87.67\hspace{3.0mm}25.90 & 61.84 \\ \hline
\multicolumn{1}{c|}{T-HuBERT} & 95.05M & LS960hr & \multicolumn{1}{c|}{61.64} & \multicolumn{1}{c|}{5.37} & {6.43} & \multicolumn{1}{c|}{3.87} & \multicolumn{1}{c|}{5.12} & \multicolumn{1}{c|}{96.27} & 0.0799 & \multicolumn{1}{c|}{98.81} & 89.27\hspace{3.0mm}22.72 & 65.28 \\ \hline\hline
\end{tabular}}
\label{tab_superb}
\vspace{-0.4cm}
\end{table*}





\vspace{-0.2cm}
\section{Results and Analysis}
\label{ssec:res_analysis}
\subsection{Main Results}
To demonstrate the effectiveness of our proposed T-HuBERT model, we first evaluate it on the standard LibriSpeech test sets.
We mainly compare T-HuBERT with three kinds of models, including 
1) speech-only pre-training models (
i.e., wav2vec 2.0, HuBERT and WavLM), 
2) speech-text pre-training models that use paired text data (
i.e., PBERT and SpeechLM), 
3) speech-text pre-training model that uses unpaired text data (
i.e., token2vec).
Table~\ref{tab_main} shows the WER results when the model is fine-tuned on 1 hour, 10 hours, 100 hours, 360 hours, and 500 hours of labeled data, respectively.
For token2vec, we only report the results without LM, since it did not release the decoding results with LM.
Firstly, our T-HuBERT achieves state-of-the-art performance with significant improvement overcall methods on low-resource scenarios (
i.e., 1-hour and 10-hour subsets).
Secondly, on the \textit{train-clean-100} split, T-HuBERT obtains 4.8/10.0\% WER, which outperforms all pre-trained models that only use speech (
i.e.,  5.7/12.0\% of WavLM) and unpaired speech and text data (
i.e., 5.1/11.8\% of token2vec).
This indicates that our T-HuBERT can better leverage additional text data in an unsupervised manner than token2vec.
However, our T-HuBERT (3.1/7.0\%) falls behind SpeechLM (2.7/6.2\%) on the \textit{train-clean-100} subset, which is due to that SpeechLM uses 100 hours of paired data for pre-training, 
while T-HuBERT does not involve any paired data during the whole training process.

To validate the effectiveness of T-HuBERT on the larger datasets, we further conduct fine-tuning experiments using \textit{train-clean-360} and \textit{train-other-500}.
The results presented at the bottom of Table~\ref{tab_main} show that our proposed T-HuBERT also outperforms the baseline HuBERT by a large margin.




\begin{table}
\centering
\caption{Performance comparison when the derived GAN-based codes are used as the targets of the last layer. The model is fine-tuned on the train-clean-100 subset.}
\scalebox{0.85}{
\begin{tabular}{c c c c c c c } 
\hline\hline
\multirow{2}{*}{Model} &\multicolumn{2}{c}{No LM} & \multicolumn{2}{c}{LM}  \\
& clean & other &  clean & other   \\
\hline
wav2vec 2.0  & 6.1  & 13.3  & 3.4  & 8.0 \\         
HuBERT      & 6.3  & 13.2  & 3.4  & 8.1 \\
token2vec   & 5.1  & 11.8  & -    &- \\
\hline
T-HuBERT (only phoneme)  & 5.0  & 11.2  & 3.1  & 7.4 \\
T-HuBERT               & 4.8  & 10.0  & 3.1  & 7.0 \\
\hline\hline
\end{tabular}
\label{tab_phn}}
\vspace{-0.3cm}
\end{table}


\subsection{Ablation Study}
\label{ssec:ablation}
\subsubsection{The effectiveness of the GAN-based tokenizer}
To verify whether the derived phoneme-like pseudo codes through GAN networks are good teachers for speech pre-training, we replace the targets of original HuBERT with our GAN-based pseudo codes and pre-train it using 960h LibriSpeech.
Table~\ref{tab_phn} shows the results when the model is fine-tuned on the \textit{train-clean-100} subset.
Obviously, when we only use the codes derived from GAN for pre-training, our T-HuBERT (5.0/11.2\%) outperforms HuBERT (6.3/13.2\%) with a large margin, and yields comparable results with token2vec (5.1/11.8\%).
It suggests the effectiveness of our proposed GAN-based tokenizer that can effectively leverage additional unpaired text in an unsupervised way to provide better targets for speech pre-training.

\subsubsection{The choice of supervision layer for the GAN-based pseudo codes}
Here we study the effect of layer choices for the GAN-based pseudo codes.
The HuBERT-based codes are always put at the 12th layer of the model, and we only vary the position of the GAN-based codes.
Due to limited resources, we conduct experiments using smaller models with embedding size 384, 6 attention heads, and feed-forward dimension 1536 in each transformer block.
As shown in Table~\ref{tab_suplayer}, the model that uses the 7th layer for the GAN-based codes yields the best performance.
When putting the GAN-based codes at a higher layer of the model, the performance starts to drop.


\subsubsection{The ratio between unpaired speech and text}
To investigate the impact of the amount of unpaired text, we gradually reduce the unpaired text used during the adversarial training of GAN and keep the speech data fixed.
From Table~\ref{tab_text}, it is obvious that the performance drops dramatically as we reduce the amount of text, which suggests that we need large amount of additional unpaired text to provide better statistics of the phoneme distribution.

\begin{table}
\centering
\caption{Results comparisons with GAN-based pseudo codes on different intermediate layers.}
\scalebox{0.9}{
\begin{tabular}{c c c c} 
\hline\hline
Phoneme layer & Target layers & test-clean & test-other\\
\hline
None  & 12   & 8.4 & 18.8 \\ \hline
6     & 6,12 & 6.6 & 14.3 \\
7     & 7,12 & 6.4 & 14.3\\
8     & 8,12 & 6.5  & 14.5\\ 
9     & 9,12 & 6.9  & 14.9\\
\hline\hline
\end{tabular}}
\label{tab_suplayer}
\vspace{-0.2cm}
\end{table}

\begin{table}[t]
\centering
\renewcommand{\arraystretch}{1.2}
\caption{WER comparison when varying the ratio between speech and text during GAN training. The model is fine-tuned on 1-hour label data of Libri-light.}
\scalebox{0.95}{
\begin{tabular}{cc}
\hline \hline
Ratio (Speech/Text) &  test-other \\ \hline
1:110               & 22.2       \\
1:50                & 29.7       \\
1:1                 & 31.6          \\ \hline 
1:0 (speech-only)   & 27.5  \\ \hline \hline
\end{tabular}}
\label{tab_text}
\vspace{-0.1cm}
\end{table}

\vspace{-0.15cm}
\subsection{Universal Speech Representation Evaluation}
\label{ssec:superb}
We further evaluate our T-HuBERT on SUPERB\cite{yang2021superb} benchmark, which provides a standard and comprehensive test-bed for evaluating the generalisability of speech pre-trained models on various tasks,
including Speaker Identification (SID), Automatic Speaker Verification (ASV), Speaker Diarization (SD), Phoneme Recognition (PR), Automatic Speech Recognition (ASR), Keyword Spotting (KS), Query by Example Spoken Term Detection (QbE), Intent Classification (IC), Slot Filling (SF), and Emotion Recognition (ER). These tasks can be grouped into four aspects of speech: content, speaker, semantics, and paralinguistics (ParaL).
We use the default fine-tuning strategy in SUPERB, which collects multiple hidden states from the pre-trained model and weighted-sum them as the final representation.
Table~\ref{tab_superb} compares T-HuBERT with other pre-training models on SUPERB.
We observe that T-HuBERT achieves better performance on semantic and paralinguistic related tasks, while on speaker related tasks, T-HuBERT also obtains competitive results.


\section{Conclusion}
\label{sec:conclusion}
This paper proposes a novel pre-training method, T-HuBERT, by leveraging additional unpaired text to improve its learned speech representations.
We build a bridge between unpaired speech and text via the generative adversarial networks in an unsupervised way and encourage the pre-trained model to capture more phonetic information.
Extensive experiments and analysis show that our model outperforms various competitive baselines on LibriSpeech and SUPERB benchmark. 
For future work, we will further improve the capabilities of pre-trained speech models with the help of large language models that contain rich semantic information.

\clearpage
\bibliographystyle{IEEEtran}
\bibliography{main}
\end{document}